\documentclass[12pt]{article}
\pdfoutput=1
\usepackage{amsbsy}
\usepackage{amsfonts}
\usepackage{amsmath,amssymb}
\usepackage{bbold}
\usepackage{pdfpages}

\newcommand{\sect}[1]{\setcounter{equation}{0}\section{#1}}

\newcommand{\eq}{\begin{equation}}
\newcommand{\eqa}{\begin{eqnarray}}  
\newcommand{\en}{\end{equation}}
\newcommand{\ena}{\end{eqnarray}}
\newcommand{\enn}{\nonumber \end{equation}}

\def\sk{\vskip .4cm}
\def\noi{\noindent}
\def\om{\omega}

\let \part\partial

\def\epsi{\varepsilon}

\def\part{\partial}

\def\sk{\vskip .4cm}

\def\noi{\noindent}

\def\X0{X^0}

\def\om{\omega}

\def\epsi{\varepsilon}

\def\Dcal{{\cal D}}

\def\Mcal{{\cal M}}
\def\Scal{{\cal S}}
\def\Gcal{{\cal G}}

\def\square{{\,\lower0.9pt\vbox{\hrule \hbox{\vrule height 0.2 cm
\hskip 0.2 cm \vrule height 0.2 cm}\hrule}\,}}

\def\Gbb{\mathbb{G}}

\def\dright{\stackrel{\rightarrow}{\part}}
\def\dleft{\stackrel{\leftarrow}{\part}}

\begin{document}

\begin{titlepage}
\rightline{ARC-2019-06}
\vskip 2em
\begin{center}
{\Large \bf Covariant hamiltonian for gravity coupled to $p$-forms} \\[3em]

\vskip 0.5cm

{\bf
Leonardo Castellani$^{1,2,3}$ and Alessandro D'Adda$^{2,3}$}
\medskip

\vskip 0.5cm

{\sl $^1$ Dipartimento di Scienze e Innovazione Tecnologica
\\Universit\`a del Piemonte Orientale, viale T. Michel 11, 15121 Alessandria, Italy\\ [.5em] $^2$ INFN, Sezione di 
Torino, via P. Giuria 1, 10125 Torino, Italy\\ [.5em]
$^3$ Arnold-Regge Center, via P. Giuria 1, 10125 Torino, Italy
}\\ [4em]
\end{center}

\begin{abstract}
\sk

We review the covariant canonical formalism initiated by D'Adda, Nelson and Regge in 1985, and extend it to include 
a definition of form-Poisson brackets (FPB) for geometric theories coupled to $p$-forms, gauging free differential algebras. The form-Legendre transformation and the form-Hamilton equations are derived from a $d$-form Lagrangian with $p$-form 
dynamical fields $\phi$.  Momenta are defined as derivatives of the Lagrangian with respect to the ``velocities" $d\phi$
and no preferred time direction is used.  Action invariance under infinitesimal form-canonical transformations can be studied in this framework, and a generalized Noether theorem is derived, both for global and local symmetries. 

We apply the
formalism to vielbein gravity in $d=3$ and $d=4$. In the $d=3$ theory we can define form-Dirac brackets, and
use an algorithmic procedure to construct the canonical generators for local Lorentz rotations 
and diffeomorphisms. In $d=4$ the canonical analysis is carried out using FPB, since the definition of form-Dirac brackets is problematic.
Lorentz generators are constructed, while diffeomorphisms are generated by the Lie derivative.

A ``doubly covariant" hamiltonian formalism is presented, allowing to maintain manifest Lorentz covariance at every stage of the 
Legendre transformation. The idea is to take curvatures as ``velocities" in the definition of momenta.

\end{abstract}

\vskip 1cm
 \noi \hrule \vskip .2cm \noi {\small
leonardo.castellani@uniupo.it \\ alessandro.dadda@to.infn.it}

\end{titlepage}

\newpage
\setcounter{page}{1}

\tableofcontents

\vfill\eject

\sect{Introduction}

Geometric theories like gravity or supergravity are conveniently formulated in the language of differential forms. The Lagrangian of a $d$-dimensional theory being written as a $d$-form, it is invariant by construction under diffeomorphisms (up to a total derivative).  This framework
is well suited also to the case of $p$-form fields coupled to (super)gravity, and a group-geometric approach has been developed since the
late 70's based on free differential algebras (FDA's) \cite{gm11}-\cite{gm23} (for a recent review see for ex. \cite{LC2018} ). 
In the 80's a form-Hamiltonian formalism was proposed in a series of papers \cite{CCF1}-\cite{CCF5}, where momenta $\pi$ conjugated to basic $p$-form fields $\phi$ are defined as ``derivatives" of the $d$-form Lagrangian with respect to the ``velocities" $d\phi$, and
the $d$-form Hamiltonian is defined as $H = (d\phi) \pi - L$. This form-Hamiltonian setting is {\it covariant}, since no preferred
(time) direction is used to define momenta. 

Other covariant Hamiltonian formalisms have been proposed in the literature, and 
a very partial list of references on multimomentum and multisymplectic canonical frameworks is given in \cite{DeDonder} -
\cite{Kaminaga2018}. The essential ideas appeared in papers by De Donder and Weyl more than seventy years ago  \cite{DeDonder,Weyl}. Some of these approaches are quite similar in spirit to the one we discuss here, but to our knowledge the first proposal of a $d$-form Hamiltonian, together with its application to gravity, can be found in ref. \cite{CCF1}. 

In this paper we further develop the form-Hamiltonian approach of ref.s \cite{CCF1}-\cite{CCF5}, and derive the Hamilton equations
for all $p$-form degrees of freedom. The form-Legendre transformation is discussed in detail, keeping track of
all necessary signs due to the presence of forms of various degrees. A definition of form-Poisson brackets (FPB) is introduced, and 
generalizes the usual Poisson brackets to arbitrary $p$-forms. These FPB satisfy generalized Jacobi identities, (anti)symmetry and derivation
properties, with signs depending on the form degrees. In this language we discuss infinitesimal canonical transformations and generators. A form-Noether theorem is derived, both for global and for local invariances of the action. 

We apply the formalism to $d=4$ tetrad gravity, and complete the analysis of \cite{CCF1,CCF2} by constructing the 
(hamiltonian) Lorentz gauge generators, acting on the basic fields via Poisson brackets. Diffeomorphisms are discussed,
and expressed in the hamiltonian setting by means of the Lie derivative. 

Vielbein gravity in $d=3$ is reformulated in the covariant hamiltonian framework, and with the use of form-Dirac brackets
we find the canonical generators for local Lorentz rotations and diffeomorphisms.

Finally, we discuss a ``doubly covariant" hamiltonian formalism for gravity (possibly coupled to $p$-forms), 
where the ``velocities" $d\phi$ are replaced by their
covariant version, i.e. the curvatures $R$. Momenta are then defined as the derivatives of $L$ with respect to $R$, and all
formulae (e.g. the Hamilton equations of motion) become automatically Lorentz covariant, derivatives being replaced throughout by
covariant derivatives.

\sect{Variational principle for geometric theories with $p$-forms}

We consider geometrical theories in $d$ dimensions with a collection of dynamical fields $\phi_i$ that are $p_i$-forms. The action $S$ is an
integral on a manifold $\Mcal^d$ of a $d$-form Lagrangian $L$ that depends on $\phi_i$ and $d \phi_i$:
\eq
S= \int_{\Mcal^d} L (\phi_i, d\phi_i)
\en
The variational principle yields 
\eq
\delta S =  \int_{\Mcal^d} \delta \phi_i { \dright L \over \partial \phi_i} + d (\delta \phi_i ) { \dright L \over \partial (d\phi_i)}=0
\en
All products are exterior products between forms. The symbol ${ \dright L \over \partial \phi_i}$ indicates the right derivative of $L$ 
 with respect to a $p$-form $\phi_i$, defined by first bringing $\phi_i$ to the left in $L$
(taking into account the sign changes due to the gradings)
and then canceling it against the derivative. In other words, we use the graded
Leibniz rule, considering ${\partial \over \partial \phi_i}$ to have the same grading as $\phi_i$. Integrating by parts\footnote{with trivial
boundary of $\Mcal^d$, or appropriate boundary
conditions.}, and since the
$\delta \phi_i$ variations are arbitrary, we find the Euler-Lagrange equations:
\eq
d ~ { \dright L \over \partial (d\phi_i)} - (-)^{p_i} { \dright L \over \partial \phi_i} =0 \label{ELeqs}
\en

\sect{Form Hamiltonian}

Here we further develop a covariant hamiltonian formalism well-adapted to geometrical theories, initiated in ref.s \cite{CCF1,CCF2,CCF3,CCF4,CCF5}. We start by defining the ($d-p_i-1$)-form momenta:
\eq
\pi^i \equiv {\dright L \over \partial (d\phi_i)} \label{momentadef}
\en
and a $d$-form Hamiltonian density (sum on $i$):
\eq
H \equiv d\phi_i ~\pi^i  - L \label{formH}
\en
This Hamiltonian density does not depend on the ``velocities" $d \phi_i$ since
\eq
{\dright H \over \partial (d\phi_i)}  = \pi^i - {\dright L \over \partial (d\phi_i)}= 0 
\en
Thus $H$ depends on the $\phi_i$ and $\pi^i$:
\eq
H=H(\phi_i,\pi^i)
\en
and the form-analogue of the Hamilton equations reads:
\eq
d \phi_i = (-)^{(d+1)(p_i+1)} {\dright H \over \partial \pi^i} ,~~~d \pi^i =  (-)^{p_i+1} ~{\dright H \over \partial \phi_i} \label{formHE}
\en
The first equation is equivalent to the momentum definition, and is obtained by taking the right derivative of $H$ as given in 
(\ref{formH}) with respect to $\pi^i$:
\eq
{\dright H \over \partial \pi^i} = {\dright d\phi_j \over \partial \pi^i} ~\pi^j + (-)^{(d-p_i - 1)(p_i+1)} d\phi_i - 
 {\dright d\phi_j \over \partial \pi^i} ~ {\dright L \over \partial (d\phi_j)}
\en
and then using (\ref{momentadef}), and $(d-p_i-1)(p_i+1) = (d+1)(p_i + 1) (mod~ 2)$.

The second is equivalent to the Euler-Lagrange form equations since 
\eq
{\dright H \over \partial \phi_i} = {\dright d\phi_j \over \partial \phi_i} ~\pi^j - { \dright L \over \partial \phi_i} -  {\dright d\phi_j \over \partial \phi_i}~ {\dright L \over \partial (d\phi_j)} = - { \dright L \over \partial \phi_i} 
\en
because of the momenta definitions (\ref{momentadef}). Then using (\ref{ELeqs}) yields the form Hamilton equation for $d\pi^i$.

\sect{Exterior differential and form Poisson bracket}

The form Hamilton equations allow to express the (on shell) exterior differential of any $p$-form $F(\phi_i, \pi^i)$ as
\eq
dF=d\phi_i ~{\dright F\over \partial \phi_i} + d \pi^i ~{\dright F\over \partial \pi^i} =  (-)^{(d+1)(p_i+1)} {\dright H \over \partial \pi^i}~
{\dright F\over \partial \phi_i} +  (-)^{p_i+1} ~{\dright H \over \partial \phi_i} {\dright F\over \partial \pi^i} 
\en
Using left derivatives this expression simplifies:
\eq
dF= {\dleft H \over \partial \pi^i}~
{\dright F\over \partial \phi_i} -  (-)^{p_i d} ~{\dleft H \over \partial \phi_i} {\dright F\over \partial \pi^i}  \label{differential}
\en
{\bf Note:} left derivatives are defined as ``acting on the left" and for example ${\dleft H \over \partial \phi_i}$ really means
${H \dleft \over \partial \phi_i}$. It is easy to verify\footnote{suppose that $A$ is contained in $F$ as $F= F_1 A F_2$. Then
${\dright F\over \partial A}  = (-)^{af_1} F_1 F_2$ and ${\dleft F\over \partial A}  = (-)^{af_2} F_1 F_2$ so that ${\dleft F\over \partial A} =
(-)^{a(f_1+f_2)} {\dright F\over \partial A} = (-)^{a(f-a)} {\dright F\over \partial A} $ and (\ref{leftright}) follows.}
that the left and right derivatives of an $f$-form $F$ with respect 
to an $a$-form $A$ satisfy
\eq
{\dleft F \over \partial A} = (-)^{a(f+1)} ~{\dright F \over \partial A} \label{leftright}
\en
and this relation is used to prove eq. (\ref{differential}).
\sk

The expression for the differential (\ref{differential}) suggests the definition of the {\it form Poisson bracket} (FPB):
\eq
\{ A, B \} \equiv  {\dleft B \over \partial \pi^i}~
{\dright A\over \partial \phi_i} -  (-)^{p_i d} ~{\dleft B \over \partial \phi_i} {\dright A\over \partial \pi^i}  \label{FPB}
\en
so that
\eq
dF = \{ F,H \} \label{differential2}
\en
{\bf Note 1:} The form Poisson bracket between the $a$-form $A$ and the $b$-form $B$ is a ($a+b-d+1$)-form, and canonically conjugated forms satisy:
\eq
 \{ \phi_i, \pi^j\}  = \delta_i^j \label{canonicalPB}
 \en
\noi {\bf Note 2:} a different definition of form Poisson bracket was given in ref. \cite{CCF1}, based on postulated properties of the FPB
rather than on the Legendre transformation that leads to the evolution equation (\ref{differential2}). In fact the properties of the FPB in 
\cite{CCF1} differ from the ones given in next Section, {\it deduced} from the definition (\ref{FPB}). 

\sect{Properties of the form Poisson bracket}

Using the definition (\ref{FPB}), the following relations can be shown to hold:
\eqa
& & \{ B,A \} = - (-)^{(a+d+1)(b+d+1)} \{ A,B \}  \label{prop1} \\
& & \{A,BC \} = B \{A,C \} + (-)^{c(a+d+1)} \{A,B \} C \\
& & \{AB,C \} =  \{A,C \} B + (-)^{a(c+d+1)}  A \{B,C \}  \\
& & (-)^{(a+d+1)(c+d+1)} \{ A, \{ B,C \} \} + cyclic~=0\\
& & (-)^{(a+d+1)(b+d+1)} \{  \{ B,C \},A \} + cyclic~=0 \label{prop5}
\ena
i.e. graded antisymmetry, derivation property, and form-Jacobi identities.

\sect{Infinitesimal canonical transformations}

We can define the action of infinitesimal form-canonical transformations on any $a$-form $A$ as follows:
\eq
\delta A = \epsi \{A,G \}
\en
where $G$ is a $(d-1)$-form, the generator of the canonical transformation, and $\epsi$ an infinitesimal parameter depending
only on the $\Mcal^d$ coordinates. Then $ \{A,G \}$ is a $a$-form like $A$.
We now prove that these transformations preserve the canonical FPB relations (\ref{canonicalPB}), thus deserving the name
of form-canonical transformations. As in the usual case the proof involves the Jacobi identities applied to $\phi_i, \pi^j, G$:
\eq
\{ \{ \phi_i, \pi^j \}, G \} + (-)^{p_i (p_i + d+1)} ~\{ \{ \pi^j , G\}, \phi_i \} + \{  \{G, \phi_i \}, \pi^j  \} =0
\en
Using the graded antisymmetry of the FPB this reduces to:
\eq
 \{\phi_i ,  \{ \pi^j , G \} \} +  \{ \{\phi_i , G \} , \pi^j \}  =  \{ \{ \phi_i, \pi^j \}, G \} =  0
\en
since $\{ \phi_i, \pi^j\}  = \delta_i^j$ is a number. Then
\eqa
& & \{\phi'_i, \pi'^j \} = \{\phi_i + \epsi  \{ \phi_i, G \}, \pi^j + \epsi  \{ \pi^j , G \} \}  \nonumber \\
& & ~~~~~~~~~~~ = \{\phi_i , \pi^j  \}  +  \epsi  \{\phi_i ,  \{ \pi^j , G \} \} +  \epsi \{ \{\phi_i , G \} , \pi^j \}  + O(\epsi^2) \nonumber \\
& & ~~~~~~~~~~~ = \{\phi_i , \pi^j  \}  + O(\epsi^2)
\ena
Q.E.D.

\sect{Form-canonical algebras}

The commutator of two infinitesimal canonical transformations generated by the ($d-1$)-forms $G_1$ and $G_2$ is again 
an infinitesimal canonical transformation, generated by the ($d-1$)-form  $\{ G_1,G_2 \}$. This is due to
\eq
\{ G_1,G_2 \} = - \{ G_2,G_1 \} 
\en
for ($d-1$)-form entries, and the form-Jacobi identity
\eq
\{ \{ A,G_1 \}, G_2 \} - \{ \{ A,G_2 \}, G_1 \} = \{ A, \{ G_1,G_2 \} \}
\en
holding for any $p$-form $A$. Therefore the form-canonical transformations close an algebra.
This algebra is finite dimensional if all fundamental fields (``positions and momenta") are $p$-forms
with $p \ge 1$, since there is only a finite number of $(d-1)$-form polynomials made out of the fundamental fields.
On the other hand, if there are fundamental $0$-forms, the algebra becomes infinite dimensional because there
are infinitely many $(d-1)$-form polynomials. 

Consider as an example a collection of 1-form fundamental fields $\phi_i$ ($i=1,...n$) in $d=4$. Their conjugated
momenta are 2-form fields $\pi^i$. There are only two types of 3-form polynomials in these fields:
\eq
G_{ijk} = \phi_i \phi_j \phi_k,~~~G_i^{j} = \phi_i \pi^j
\en
Their (finite) Poisson bracket algebra reads
\eq
\{ G_{ijk}, G_{lmn} \} =0,~~~\{ G_{ijk}, G_l^m \} = 3 \delta_{[k}^m ~ G_{ij]l},~~~ \{ G_i^j, G_k^l \} = \delta_i^l G_k^j - \delta_k^j G_i^l
\en
with $m$ = ${n}\choose{3}$  generators $G_{ijk}$  closing on a $U(1)^m$ subalgebra and $n^2$ generators $G_i^{j}$ closing on a $U(n)$ subalgebra. 
The whole algebra is then a semidirect sum of $U(n)$ with $U(1)^m$ . 

\sect{Action invariance and Noether theorem} 

\subsection{Global invariances}

Consider the action 
\eq
S=\int_{\Mcal^d} d \phi_i~\pi^i - H
\en
Its variation under an infinitesimal form-canonical transformation generated by a ($d-1$)-form $G$ is
\eqa
& & \delta S = \int_{\Mcal^d} d (\{ \phi_i , G \} ) \pi^i + d \phi_i \{ \pi^i , G \} - \{ H, G \} \nonumber \\
& & ~~~~ =    \int_{\Mcal^d} d (\{ \phi_i , G \}  \pi^i ) + (-)^{p_i+1}  \{ \phi_i , G \}  d\pi^i  + d \phi_i \{ \pi^i , G \} - \{ H, G \} \nonumber \\
& & ~~~~ =    \int_{\Mcal^d} d (\{ \phi_i , G \}  \pi^i ) + (-)^{p_i+1}  ~{\dleft G \over \partial \pi^i} ~ d \pi^i - (-)^{p_i d} ~d \phi_i ~
 {\dleft G \over \partial \phi_i} - \{ H, G \} \nonumber \\
& & ~~~~ =    \int_{\Mcal^d} d (\{ \phi_i , G \}  \pi^i ) + (-)^{p_i+1} (-)^{p_i} ~ d \pi^i~ {\dright G \over \partial \pi^i} - (-)^{p_i d} (-)^{p_i d} ~d \phi_i ~ {\dright G \over \partial \phi_i} - \{ H, G \} \nonumber \\
 & & ~~~~ =   \int_{\Mcal^d} d (\{ \phi_i , G \}  \pi^i ) - d \pi^i~ {\dright G \over \partial \pi^i} - d \phi_i ~ {\dright G \over \partial \phi_i} - \{ H, G \} \nonumber \\
  & & ~~~~ =   \int_{\Mcal^d} d (\{ \phi_i , G \}  \pi^i ) - d G  - \{ H, G \} \nonumber \\
  & & ~~~~ =   \int_{\partial \Mcal^d}  (\{ \phi_i , G \}  \pi^i  -  G  ) -  \int_{\Mcal^d} \{ H, G \}  \label{Noether1}
\ena
Thus the action is invariant (up to a boundary term) under the infinitesimal canonical form-transformation generated by $G$ iff 
\eq
\{ H, G \} =0
\en
up to a total derivative. This result reproduces Noether's theorem in form language.  
\sk
\noi {\bf Note: } here $G$ is a polynomial in the $\phi_i$ and $\pi^i$. In this case 
\eq
dG =  d \pi^i~ {\dright G \over \partial \pi^i} + d \phi_i ~ {\dright G \over \partial \phi_i} 
\en
has been used in the sixth line of (\ref{Noether1}). Generators containing spacetime functions $f(x)$ (``external fields") are considered in the next paragraph.
\sk
\noi On shell we have
\eq
dG = \{ G, H \} 
\en
Thus if $G$ generates an invariance of the action, on shell its exterior derivative vanishes. Consider then
the $d$-dimensional integral
\eq
\int dG
\en
between two ($d-1$)-dimensional spacelike slices $\Scal_{t_1}$ and $\Scal_{t_2}$ of the $\Mcal^d$ manifold corresponding to the times $t_1$ and $t_2$.  By Stokes theorem this integral is equal to the difference between the integrals of $G$ on the $\Scal_{t_2}$ and $\Scal_{t_1}$ slices, and since $dG=0$, this difference vanishes, implying that the $0$-form quantity
\eq
\Gcal (t) = \int_{\Scal_t} G
\en
is conserved in time on the shell of the equations of motion\footnote{If $\{ H, G \} = dW$, then $d(G-W) =0$ on shell
and $\int_{\Scal_t} G-W$ is conserved in time.}.

\subsection{Gauge invariances generated by $\epsi (x) G$}

Here we consider generators of the type $\epsi (x) G$, generating $x$-dependent infinitesimal form-canonical 
transformations:
\eq
\delta \phi_i = \epsi (x) \{ \phi_i, G \},~~~\delta \pi^i = \epsi (x) \{ \pi^i, G \}
\en
The variation of the action is computed along the same lines of the preceding subsection, with an additional term
due to the infinitesimal parameter $\epsi$ being nonconstant, and reads:
\eq
\delta S =   \int_{\partial \Mcal^d} \epsi (\{ \phi_i , G \}  \pi^i  -G)  +  \int_{\Mcal^d} (d\epsi  ~G - \epsi \{ H, G \} )  \label{Noether2}
\en
Thus $\epsi(x) G$ is a gauge generator, leaving the action invariant (up to boundary terms) iff
\eq
G=0,~~~ \{H,G \} =0
\en
\noi since $\epsi(x)$ is an arbitrary function. Thus $G$ and  $\{H,G \}$ must be {\it  constraints}.

If there is a collection of ($d-1$)-forms $G_A$ generating local invariances of the action\footnote{here and in the following, invariance of the action will be understood up to surface terms.}, the commutator of two
transformations generated by $G_1$ and $G_2$ must leave the action invariant. This commutator is generated by 
$\{ G_1, G_2 \}$ because of Jacobi identities. Therefore $\{G_A, G_B \} $ is a gauge generator. The gauge algebra
can involve structure constants
\eq
\{G_A,G_B \} = C^C_{AB} ~G_C
\en
as in ordinary finite Lie algebras, or structure functions, as is the case of diffeomorphisms 
in gravity theories. 

Finally, the infinitesimal transformations generated by $\epsi(x) G$ must preserve the constraints, and therefore
\eq
\{ constraints, G \} \approx 0
\en
\noi where $\approx$ means weak equality, i.e. holding on the constraint surface.

\subsection{Gauge invariances generated by $\epsi (x) G + (d \epsi) F $}

In gauge and gravity theories the infinitesimal symmetry transformations on the fields contain also
derivatives of the $x$-dependent parameter. We need thus to consider generators of the form
$\epsi (x) G + (d \epsi) F $, where $F$ is a ($d-2$)-form, and investigate how they transform the action. The answer is 
\eqa
& & \delta S =   \int_{\partial \Mcal^d}  \epsi (\{ \phi_i , G \}  \pi^i  -  G  ) + d\epsi (\{ \phi_i , F \}  \pi^i  -  F ) \nonumber \\
& & ~~~~~~ +  \int_{\Mcal^d} [d \epsi (G - \{H,F \}) -  \epsi \{ H, G \}]  \label{Noether3}
\ena
Thus $\epsi (x) G + (d \epsi) F $ is a gauge generator leaving the action invariant iff
\eq
G -\{H,F \} =0,~~~\{H,G \} =0 \label{conditions2}
\en
Moreover the infinitesimal transformation generated by $\epsi (x) G + (d \epsi) F $ must preserve the constraints, implying
\eq
\{ constraints, G \} \approx 0,~~~\{ constraints, F \} \approx 0  \label{conditions2bis}
\en
The conditions (\ref{conditions2}) and (\ref{conditions2bis}) generalize to geometric theories with fundamental
$p$-form fields the conditions for gauge generators found in \cite{SCHS}, and provide the basis for a constructive
algorithm yielding all the gauge generators. We illustrate the procedure in the next Sections.
\sk
\noi {\bf Note 1:} $F$ and $G$ must be first-class quantities, i.e. have weakly vanishing FPS's with all the constraints,
but do not have necessarily to be constraints.
\sk
\noi{\bf Note 2:} this Section reproduces the results of \cite{SCHS}, in the present context of geometric theories with 
fundamental $p$-forms.
\sk
\noi {\bf Note 3:}  in the form setting the time derivatives of usual canonical formalism
become exterior derivatives, and due to $d^2 =0$ gauge generators cannot contain second or higher derivatives of $\epsi$.
Thus geometric theories do not give rise to tertiary constraints, since these would multiply second derivatives of
the gauge parameter in the gauge generator chains \cite{SCHS}.

\sect{Gravity in $d=4$}

\subsection{Form hamiltonian and constraints}

The fields $\phi_i$ in this case
are 1-forms: the vierbein $V^a$ and the spin connection $\om^{ab}$. Torsion and Lorentz curvature are defined as usual:
\eq
R^a = dV^a -\om^a_{~b} ~ V^b,~~~R^{ab}= d \om^{ab} - \om^a_{~e}~ \om^{eb} \label{curvatures}
\en
and the Einstein-Hilbert 4-form Lagrangian is
\eq
L (\phi, d\phi) = R^{ab} V^c V^d \epsi_{abcd} = d \om^{ab} V^c V^d \epsi_{abcd} - \om^a_{~e} \om^{eb} V^c V^d \epsi_{abcd}
\label{EHLagrangian}
\en
The 2-form momenta conjugated to $V^a$ and $\omega_{ab}$ are respectively\footnote{unless stated otherwise, all partial
derivatives act from the left in the following.} :
\eqa
& & \pi_{a} = {\partial L \over \partial (dV^a)} = 0 \\
& & \pi_{ab} = {\partial L \over \partial (d \om^{ab})} =V^c V^d \epsi_{abcd}
\ena
Both momenta definitions are {\it primary constraints}:
\eq
\Phi_a \equiv \pi_a = 0,~~~\Phi_{ab} \equiv \pi_{ab} - V^cV^d \epsi_{abcd} = 0
\en
 since they do not involve the ``velocities" $dV^a$ and $d\om^{ab}$. 
The form Hamiltonian is:
\eqa
& & H= dV^a ~ \pi_a + d \om^{ab}~ \pi_{ab} - d \om^{ab} ~V^c V^d \epsi_{abcd} + \om^a_{~e} ~\om^{eb}~ V^c  V^d \epsi_{abcd}= \nonumber \\
& & ~~~ = dV^a ~ \Phi_a + d \om^{ab}~ \Phi_{ab}  + \om^a_{~e} ~\om^{eb}~ V^c V^d  \epsi_{abcd}
\ena
The ``velocities"  $dV^a$ and
$d\om^{ab}$ are undetermined at this stage. Indeed the Hamilton equations of motion for $dV^a$ and $d\omega^{ab}$
are just identities ($dV^a=dV^a$, $d\omega^{ab}=d\omega^{ab}$), whereas for the momenta they read:
\eqa
& & d \pi_a =  {\partial H \over \partial V^a} = - 2 R^{bc} V^d \epsilon_{abcd} \\
& & d\pi_{ab} =  {\partial H \over \partial \omega^{ab}} = 2 \omega^c_{~[a} V^d V^e \epsilon_{b]cde}
\ena

Requiring the ``conservation" of
$\Phi_a$ and $\Phi_{ab}$, i.e. their {\it closure} in the present formalism, leads to the
conditions:
\eqa
& & d \Phi_a = \{ \Phi_a,H \} =0 ~~~\Rightarrow ~~~ R^{bc} ~V^d \epsi_{abcd}  = 0 \label{secondary1} \\
& & d \Phi_{ab} = \{ \Phi_{ab},H \} =0 ~~\Rightarrow ~~~  R^c ~V^d \epsi_{abcd} = 0 \label{secondary2}
\ena
To derive (\ref{secondary2}) we also made use of the identity
\eq
F^e_{~[a} \epsi_{bcd]e} =0 \label{identity1}
\en
holding for any antisymmetric $F$. 
The conditions (\ref{secondary1}), (\ref{secondary2}) are respectively equivalent to the Einstein field equations
and to the zero torsion condition $R^a =0$, that enables to express the spin connection in terms of the vierbein.  
Note that we cannot call them secondary constraints, since they contain the ``velocities" $dV^a$ and $d\omega^{ab}$.
In fact, they {determine  $dV^a$ as
\eq
dV^a = \om^a_{~b}~ V^b
\en
and determine some (combinantions of) components of $d \omega^{ab}$ by constraining $R^{ab}$
via the Einstein equations.
 
Using the form bracket we find the constraint algebra:
  \eq
  \{ \Phi_a,\Phi_b \}=  \{ \Phi_{ab},\Phi_{cd} \}=0;~~~  \{ \Phi_a,\Phi_{bc} \}=-2\epsi_{abcd} V^d
  \en
 showing that the constraints are not all first-class. This is consistent with the fact that some of the undetermined ``velocities" 
get fixed by requiring conservation of the primary constraints. Classical references on constrained hamiltonian systems are given in
\cite{Dirac,HRT,HT}.
\sk
\noi {\bf Note:} the action variations  (\ref{Noether2}) and (\ref{Noether3}) have been deduced assuming that $H$ depends only on
basic fields and momenta. This is not the case in constrained systems, where some of the velocities remain undetermined,
and therefore appear in the hamiltonian. However they always appear multiplied by primary constraints, and the
variation of these terms always vanishes weakly. 

\subsection{Gauge generators}
\sk
{\bf Lorentz gauge transformations}
\sk
We start from the first class 2-forms $\pi_{ab}$, having vanishing FPB's with the constraints $\Phi_a$, $\Phi_{ab}$. They
will play the role of the ($d-2$)-forms $F$ of Section 8.3, with two antisymmetric indices, thus $F_{ab} =  \pi_{ab}$. To find the corresponding ($d-1$)-form $G_{ab}$ that complete the
gauge generators one uses the first condition in (\ref{conditions2}), yielding $G_{ab}$ as the PB of $H$ with $F_{ab}$, up to constraints.
Since
\eq
\{H,\pi_{ab} \} = 2 \omega_{[a}^{~~e} V^c V^d \epsilon_{b]ecd}
\en
we find that
\eq
G_{ab} = 2 \omega_{[a}^{~e} V^c V^d \epsilon_{b]ecd} + \alpha_{ab}^c ~ \Phi_c + \beta_{ab}^{cd} ~\Phi_{cd}
\en
\noi where $\alpha_{ab}^c $ and $\beta_{ab}^{cd} $ are 1-form coefficients to be determined by the second condition in
(\ref{conditions2}), i.e. weak vanishing of the PB between $H$ and $G_{ab}$. This yields
\eq
\alpha_{ab}^c = \delta^c_{[a} V_{b]},~~~\beta_{ab}^{cd} = 2 \omega_{[a}^{~~c}~ \delta^d_{b]}
\en
so that $G_{ab}$ becomes:
\eq
G_{ab} =  2 \omega^c_{~[a} \pi_{b] c} - V_{[a} \pi_{b]}
\en
It is easy to check that this $G_{ab}$ has weakly vanishing PB's with the constraints $\Phi_a$, $\Phi_{ab}$ and is
therefore a first-class 3-form. We have thus constructed the gauge generator
\eq
\Gbb = \epsi^{ab} G_{ab} + d \epsi^{ab} F_{ab} = \epsi^{ab} (2 \omega^c_{~a} \pi_{bc} - V_{a} \pi_{b}) + (d \epsi^{ab}) \pi_{ab} =
 \Dcal \epsi^{ab} \pi_{ab} - \epsi^{ab} V_a \pi_b
\en
It generates the Lorentz gauge rotations on all canonical variables. Indeed
\eqa
& & \delta V^a = \{V^a, \Gbb \} = \epsi^a_{~b} V^b, ~~~\delta \omega^{ab} = \{\omega^{ab}, \Gbb \} = \Dcal \epsi^{ab} \\
& &  \delta \pi_a = \{\pi_a, \Gbb \} = \epsi_a^{~b} \pi_b , ~~~~~\delta  \pi_{ab} =  \epsi_{~[a}^{c} \pi_{b] c}  
\ena
and satisfies all the conditions to be a symmetry generator of the action.

\sect{Lie derivative and diffeomorphisms}

Infinitesimal diffeomorphisms on $p$-forms $A$ are expressed by means of the Lie derivative $\ell_\epsi$:
\eq
\delta A = \ell_\epsi A \equiv ( \iota_\epsi d + d \iota_\epsi ) A
\en
where $\iota_\epsi$ is the contraction along the tangent vector $\epsi (x) = \epsi^\mu (x)  \partial_\mu$. Geometric
theories are by construction invariant under diffeomorphisms, since the action is an integral of a $d$-form on a $d$-dimensional manifold.

The variations under infinitesimal diff.s of the basic fields of $d=4$ first order tetrad gravity are
 \eqa
 & & \delta V^a =  \iota_\epsi d V^a+ d (\iota_\epsi V^a) = \Dcal \epsi^a +2 R^a_{~bc} ~\epsi^b V^c+ (\epsi^\mu \omega^{ab}_\mu) V_b \label{diffV}  \\
 & & \delta \om^{ab} =  \iota_\epsi d \omega^{ab}  + d (\iota_\epsi \omega^{ab} )= 2 R^{ab}_{~~cd} ~ \epsi^c V^d + 2 (\epsi^\mu \omega^{c[a}_\mu) \omega_{~c}^{b]}
 \ena
 where $\epsi^a \equiv \epsi^\mu V^a_\mu$, $\Dcal$ is the Lorentz covariant derivative  $\Dcal \epsi^a \equiv d \epsi^a - \omega^a_{~b} \epsi^b$, and $R^a_{bc}$ are the flat components of the torsion 2-form $R^a$, thus $R^a = R^a_{bc} V^b V^c$ and similar for the Lorentz curvature $R^{ab}$. 
 
  The infinitesimal diff.s on the momenta 2-forms are given by:
  \eqa
  & & \delta \pi_a = \iota_\epsi d \pi_a+ d (\iota_\epsi \pi_a) = \iota_\epsi (\Dcal \pi_a) + \Dcal (\iota_\epsi \pi_a) + (\epsi^\mu \omega_{a~~\mu}^{~b}) \pi_b \\
   & & \delta \pi_{ab} = \iota_\epsi d \pi_{ab}+ d (\iota_\epsi \pi_{ab}) = \iota_\epsi (\Dcal \pi_{ab}) + \Dcal (\iota_\epsi \pi_{ab}) + 2 (\epsi^\mu \omega_{~[a~\mu}^{c}) \pi_{b]c}
  \ena
  
We see that in all these variations the last term is really a Lorentz rotation with parameter 
$\eta^{ab} = \epsi^\mu \omega^{ab}_\mu$. The action being invariant under Lorentz transformations, the following
variations
\eqa
 & & \delta V^a = \Dcal \epsi^a +2 R^a_{~bc} ~\epsi^b V^c  \label{diffV} \\
 & & \delta \om^{ab} = 2 R^{ab}_{~~cd} ~ \epsi^c V^d  \label{diffomega}\\
  & & \delta \pi_a = \iota_\epsi (\Dcal \pi_a) + \Dcal (\iota_\epsi \pi_a)  \\
   & & \delta \pi_{ab} = \iota_\epsi (\Dcal \pi_{ab}) + \Dcal (\iota_\epsi \pi_{ab}) 
 \ena
generate by themselves symmetries of the action. In fact (\ref{diffV}) and (\ref{diffomega}) are the diff.s transformations deduced from the group manifold approach to first order tetrad gravity, see for ex. ref.s  \cite{gm21,LC2018}.

We may wonder whether the infinitesimal diff.s could be expressed as canonical transformations via the FPB. In the present 
form-canonical scheme this seems impossible. The reason is that the would-be generator of the diff.s, of the type
\eq
\Gbb = \epsi (x) G + (d \epsi) F 
\en
should be such that the 2-form $F$ is a first-class quantity. However there is only one such quantity, namely $\pi_{ab}$, that we
have already used in the construction of the Lorentz canonical generators. Indeed $\pi_a$ does not have weakly vanishing FPB
with the constraints $\Phi_{ab}$. We can write down a canonical generator that reproduces the correct
infinitesimal diff.s on $V^a$ and $\omega^{ab}$:
\eq
\Gbb =  \epsi^a  (2 R^b_{~ac}V^c \pi_a + 2 R^{bc}_{~~ad} V^d \pi_{bc}) + (\Dcal \epsi^a) \pi_a 
\en
but this $\Gbb$ does not generate the correct diff.s on the momenta $\pi_a,\pi_{ab}$, and does not satisfy
all the conditions of Sect. 8 for a gauge generator.

\sect{Gravity in $d=3$}

\subsection{Form hamiltonian and constraints}

The fields $\phi_i$ 
are the $d=3$ vierbein $V^a$ and the spin connection $\om^{ab}$. Torsion $R^a$ and Lorentz curvature $R^{ab}$ are defined as 
in (\ref{curvatures}),  and the Einstein-Hilbert 3-form Lagrangian is
\eq
L (\phi, d\phi) = R^{ab} V^c \epsi_{abc} = d \om^{ab} V^c \epsi_{abc} - \om^a_{~e} \om^{eb} V^c  \epsi_{abc}
\label{EHLagrangiand3}
\en
The 1-form momenta conjugated to $V^a$ and $\omega_{ab}$ are respectively :
\eqa
& & \pi_{a} = {\partial L \over \partial (dV^a)} = 0 \\
& & \pi_{ab} = {\partial L \over \partial (d \om^{ab})} =V^c  \epsi_{abc}
\ena
Both momenta definitions are {\it primary constraints}:
\eq
\Phi_a \equiv \pi_a = 0,~~~\Phi_{ab} \equiv \pi_{ab} - V^c \epsi_{abc} = 0
\en
 since they do not involve the ``velocities" $dV^a$ and $d\om^{ab}$.  The 3-form Hamiltonian is:
\eqa
& & H= dV^a ~ \pi_a + d \om^{ab}~ \pi_{ab} - d \om^{ab} ~V^c  \epsi_{abc} + \om^a_{~e} ~\om^{eb}~ V^c \epsi_{abc}= \\
& & ~~~ = dV^a ~ \Phi_a + d \om^{ab}~ \Phi_{ab}  + \om^a_{~e} ~\om^{eb}~ V^c  \epsi_{abc}
\ena
The Hamilton equations of motion for $dV^a$ and $d\omega^{ab}$
are identities,  while for the momenta they read:
\eqa
& & d \pi_a =  {\partial H \over \partial V^a} = - 2 R^{bc}  \epsilon_{abc} \\
& & d\pi_{ab} =  {\partial H \over \partial \omega^{ab}} = 2 \omega^c_{~[a} V^d  \epsilon_{b]cd}
\ena
Requiring the ``conservation" of
$\Phi_a$ and $\Phi_{ab}$ leads to the
conditions:
\eqa
& & d \Phi_a = \{ \Phi_a,H \} =0 ~~~\Rightarrow ~~~ R^{bc} \epsi_{abc}  = 0 \label{secondary1d3} \\
& & d \Phi_{ab} = \{ \Phi_{ab},H \} =0 ~~\Rightarrow ~~~  R^c \epsi_{abc} = 0 \label{secondary2d3}
\ena
implying the vanishing of both curvatures: $R^a =0$, $R^{ab}=0$. These are the equations of motion of $d=3$ 
first-order vielbein gravity. These equations completely determine the ``velocities" $dV^a$ and $d\omega^{ab}$:
\eq
dV^a=\omega^a_{~b} ~V^b,~~~d \omega^a_{~b} = \omega^a_{~c}~ \omega^{cb}
\en
 
Using the form bracket we find the constraint algebra:
  \eq
  \{ \Phi_a,\Phi_b \}=  \{ \Phi_{ab},\Phi_{cd} \}=0;~~~  \{ \Phi_a,\Phi_{bc} \}=-\epsi_{abc} 
  \en
\noi  all other FPB's vanishing. Thus constraints are second-class, and this is consistent with the fact that all the  ``velocities" 
get fixed by requiring conservation of the primary constraints. The three constraints $\Phi_{ab}$ ($ab=12,13,23$) are equivalent
to the three linear combinations $\Xi^a= {1 \over 2} \epsilon^{abc} \Phi_{bc}$, and we find
\eq
\{ \Phi_a, \Xi^b \} = \delta^b_a
\en
We'll use the $\Xi^a$ in the definition of Dirac brackets of next Section. Note that form-Poisson brackets between 1-forms are symmetric in $d=3$, and in all odd dimensions, see eq. (\ref{prop1}). Also, the FPB betwen constraints yield numbers in $d=3$ gravity, and this allows a definition of form-Dirac brackets (see next Section). A similar definition is not available in $d=4$, since the FPB between constraints yield 1-forms, and the corresponding FPB matrix has no obvious inverse. 

\subsection{Form Dirac brackets}

We define form Dirac brackets as follows
\eq
\{f,g \}^* \equiv \{f,g \} - \{f, \Phi_a \} \{ \Xi^a, g \} - \{f, \Xi^a \} \{\Phi_a,g \}
\en
These brackets vanish strongly if any entry is a constraint $\Phi_a$ or $\Xi^a$. With the help of the general formulas
(\ref{prop1})-(\ref{prop5}) with $d=3$ it is straightforward to show that the Dirac brackets inherit the same properties of the Poisson brackets,
i.e. :
\eqa
& & \{ B,A \}^* = - (-)^{ab} \{ A,B \}^*  \label{prop1d3} \\
& & \{A,BC \}^* = B \{A,C \}^* + (-)^{ca} \{A,B \}^*C \label{prop2d3}\\
& & \{AB,C \}^* =  \{A,C \}^* B + (-)^{ac}  A \{B,C \}^* \label{prop3d3} \\
& & (-)^{ac} \{ A, \{ B,C \}^* \}^* + cyclic~=0\\
& & (-)^{ab} \{  \{ B,C \}^*,A \}^* + cyclic~=0
\ena
Using Dirac brackets the second-class constraints (i.e. all the constraints of the $d=3$ theory) disappear from the game, and 
we can use the 3-form Hamiltonian
\eq
H =   \om^a_{~e} ~\om^{eb}~ V^c  \epsi_{abc}
\en
The Dirac brackets between the basic fields and their momenta are given by:
\eqa
& & \{ V^a, V^b \}^* = 0,~~~\{\omega^{ab}, \omega^{cd} \}^* =0,~~~ \{ V^a, \omega^{bc} \}^* = - {1 \over 2} \epsilon^{abc} \\
& & \{ any, \pi_a \}^* =0,~~~\{V^a, \pi_{bc} \}^* =0,~~~\{ \omega^{ab}, \pi_{cd} \}^*= \delta^{ab}_{cd}
\ena
Thus $V^a$ and $\Omega_b \equiv \epsilon_{bcd} \omega^{cd}$ become canonically conjugated variables:
\eq
\{ V^a , \Omega_b \}^*=\delta^a_b
\en
The Hamilton equations expressed via the Dirac bracket become:
\eqa
& & dV^a = \{ V^a, H \}^* = \{ V^a, \om^d_{~e} ~\om^{eb}~ V^c  \epsi_{bcd} \}^* = \omega^a_{~b} V^b~~\Rightarrow R^a=0 \\
& & d \omega^{ab}= \{ \omega^{ab}, H \}^* =  \{ \omega^{ab}, \om^d_{~e} ~\om^{ef}~ V^c  \epsi_{fcd} \}^* = \omega_e^{~[a} \omega^{b]}_{~~e} ~~\Rightarrow R^{ab}=0 \nonumber\\
\ena
i.e. the field equations of $d=3$ first order vielbein gravity. For the ``evolution" of the momenta we find:
\eqa
& & d \pi_a = \{ \pi_a, H \}^* = 0 \\
& & d \pi_{ab} =  \{ \pi_{ab}, H \}^* =  2 \omega^c_{~[a} V^d  \epsilon_{b]cd}= \epsilon_{abc} \omega^c_{~d} V^d ~~~\Rightarrow d \Phi_{ab}=0
\ena
where in the last line we used the identity 
\eq
\omega_{[a}^{~d} \epsilon_{bc]d} =0
\en
The momenta evolutions re-express the fact that the constraints are conserved, or equivalently that the exterior derivative of the momenta
is in agreement with their expression given by the second-class constraints.

\subsection{Gauge generators}

Now we apply our procedure to find the gauge generators. Here besides the Lorentz generators we will find also the canonical
generators for diffeomorphisms.
\sk
\noi {\bf Lorentz gauge transformations}
\sk
We start from the first class 1-forms $\pi_{ab}$. They are first class in the sense that they have vanishing Dirac brackets
with all the constraints. Actually the constraints being all second class, they have been effectively eliminated from the theory 
by the use of Dirac brackets.  We take these 1-forms $\pi_{ab}$ as the ($d-2$)-forms $F$  in eq. (\ref{conditions2}), and find the ($d-1$)-forms $G$ that completes the gauge generator:
\eq
G_{ab} = \{H,F_{ab} \}^* = \{ H, \pi_{ab}\}^*  =  2 \omega^c_{~[a} V^d  \epsilon_{b]cd}  \label{condition3}
\en 
Next we have to check that $\{H,G_{ab} \}=0$. Notice that here it is useless to add to $G_{ab}$ any combination of constraints, since 
second-class constraints have no effect in a generator when using Dirac brackets. So $\{H,G_{ab} \}^*=0$ must hold with the $G_{ab}$ as given in (\ref{condition3}), and indeed this is the case: the bracket yields terms  $\omega \omega V$ that sum to zero, using the $\{V,\omega\}^*$ bracket and the properties (\ref{prop2d3}), (\ref{prop3d3}).
Thus
\eq
\Gbb = d\epsilon^{ab} F_{ab} + \epsilon^{ab} G_{ab} =  d\epsilon^{ab} \pi_{ab} + 2 \epsilon^{ab}  \omega^c_{~[a} V^d  \epsilon_{b]cd}
\en
generates gauge transformations via the Dirac bracket. Using the (second-class) constraint $\pi_{ab}= \epsilon_{abc} V^c$ in the
second term of the generator yields
\eq
\Gbb=d\epsilon^{ab} \pi_{ab} + 2 \epsilon^{ab}  \omega^c_{~[a} \pi_{b]c}=  (\Dcal \epsi^{ab} ) \pi_{ab}
\en
It generates local Lorentz transformations with parameter $\epsilon_{ab} (x)$, since
\eqa
& & \delta V^a = \{V^a, \Gbb \}^* = 2 \{\omega^{[b}_{~d}, V^a \}^* \epsilon^{c]d} \pi_{bc} = \epsilon^a_{~b} V^b \\
& & \delta \omega^{ab} = \{ \omega^{ab}, \Gbb \}^* = \Dcal \epsi^{ab} \\
& & \delta \pi_a = \{ \pi_a, \Gbb \}^* =0 \\
& &  \delta \pi_{ab} = \{ \pi_{ab}, \Gbb \}^* =  \{ \epsilon_{abc} V^c, \Gbb \}^* =  \epsi_{~[a}^{c} \pi_{b] c} 
\ena
Note that  $\delta \pi_a = 0$ since $\Gbb$ has no effect on second class constraints.
\sk
\noi {\bf Diffeomorphisms}
\sk
The procedure of the preceding paragraph can be started with any 1-form: indeed here any 1-form has
vanishing Dirac brackets with the constraints. We choose $F_a$ to be $\epsilon_{abc} \omega^{bc}$, since
this 1-form is conjugated to $V^a$, and therefore a good candidate to multiply the $d \epsi^a$ term in 
the generator of the diffeomorphisms. Then $G_a$ is found in the usual way:
\eq
G_a =  \{H, F_a \}^* =  \epsilon_{abc} ~\omega^b_{~d} \omega^{dc}
\en
We have now to check that the second condition in (\ref{conditions2}) is satisfied, i.e. that 
\eq
\{ H, G_a \}^* = \{ H, \epsilon_{abc} ~\omega^b_{~d} \omega^{dc} \}^* = \epsilon_{abc} \omega^b_{~d} \omega^d_{~e} \omega^{ec} =0
\en
This is indeed so, as we can verify by specializing indices (for ex. choose $a=1$ and explicitly perform the sum on the
other indices. The result vanishes because in each term $\omega \omega \omega$ two $\omega$'s have always the
same indices). Therefore 
\eq
\Gbb = d \epsi^a F_a + \epsi^a G_a = (d \epsi^a) \epsilon_{abc} \omega^{bc} + \epsi^a \epsilon_{abc} ~\omega^b_{~d} \omega^{dc}
= ({\cal D} \epsi^a) \epsi_{abc} \omega^{bc}
\label{diffgenerator}
\en
generates a symmetry. Its action on the basic fields is given by:
\eqa
& & \delta V^a = \{ V^a, \Gbb \}^* = \Dcal \epsi^a  \label{d3diff1}\\
& & \delta \omega^{ab}= \{ \omega^{ab} , \Gbb \}^* = 0 \\
& & \delta \pi_a = \{ \pi_a, \Gbb \}^*=0 \\
& & \delta \pi_{ab} = \{ \pi_{ab} , \Gbb \}^* = \{\epsilon_{abc}  V^c , \Gbb \}^*  = \epsilon_{abc} \Dcal \epsi^c \label{d3diff4}
\ena
This infinitesimal transformation has to be compared with the infinitesimal diffeomorphisms discussed in Section 10.
In second order formalism, i.e. when $R^a = 0$ holds, the above transformations of $V^a$ and $\omega^{ab}$ are
indeed diffeomorphisms, since the $R^a$ term of (\ref{diffV}) vanishes, and the variation of the spin connection can be
taken equal to zero since it multiplies its own field equation when varying the action (this is the essence of the
so-called {\it 1.5 order formalism}, used to prove invariance of the $d=4$ supergravity action under local supersymmetry variations
\cite{PvN}). Since the $\omega^{ab}$ field equation is equivalent to $R^a=0$, any variation of $\omega^{ab}$ has no effect on the action when using $R^a=0$. Thus we can consider (\ref{diffgenerator}) to be the diffeomorphism generator of
$d=3$ gravity in second order formalism.
\sk
\noi {\bf Note: } invariance of the action under the transformations (\ref{d3diff1})-(\ref{d3diff4}) can be checked directly using integration by parts and the Bianchi identity ${\cal D} R^{ab}=0$.

\sect{A ``doubly covariant" hamiltonian for gravity}

Exploiting Lorentz symmetry, we can reformulate the form-canonical scheme for gravity in an even more covariant
way. We call this scheme "doubly covariant", in the sense that not only there is no preferred time direction in
the definition of form-momenta, but all tensors appearing in the Hamiltonian and the equations of motion are Lorentz
covariant tensors.

To achieve this, it is sufficient to take as ``velocities" not the exterior derivatives o $V^a$ and $\omega^{ab}$, but
their Lorentz covariant version, i.e. the curvatures $R^a$ and $R^{ab}$. The momenta are defined then as:
\eqa
& & \pi_{a} = {\partial L \over \partial R^a} = 0 \\
& & \pi_{ab} = {\partial L \over \partial R^{ab}} =V^c V^d \epsi_{abcd}
\ena
Both momenta definitions coincide with those of Sect. 9 and yield the same  primary constraints:
\eq
\Phi_a \equiv \pi_a = 0,~~~\Phi_{ab} \equiv \pi_{ab} - V^cV^d \epsi_{abcd} = 0
\en
 since they do not involve the ``velocities" $R^a$ and
$R^{ab}$. 
The doubly covariant form Hamiltonian is:
\eq
H= R^a ~ \pi_a + R^{ab}~ \pi_{ab} - R^{ab} V^c V^d \epsi_{abcd}   =  R^a ~ \pi_a + R^{ab}~ \Phi_{ab} 
\en
and is a sum of primary constraints. It differs from the Hamiltonian of Sect. 9, that was not a sum of primary constraints.
The Hamilton equations of motion are 
\eqa
& & R^a = \{ V^a, H\} = R^a \\
& & R^{ab}  =  \{ \omega^{ab}, H\} = R^{ab} \\
& & \Dcal \pi_a = \{ \pi_a, H \} = -2 R^{bc} V^d \epsilon_{abcd} \\
& &  \Dcal \pi_{ab} = \{ \pi_{ab}, H \} = 0
\ena
The FPB's here are defined as to leave untouched the ``velocities" $R^a$, $R^{ab}$. 

Requiring the ``covariant conservation" of
$\Phi_a$ and $\Phi_{ab}$ leads to the
conditions:
\eqa
& & \Dcal \Phi_a = \{ \Phi_a,H \} =0 ~~~\Rightarrow ~~~ R^{ab} ~V^d \epsi_{abcd}  = 0 \label{DCsecondary1} \\
& & \Dcal \Phi_{ab} = \{ \Phi_{ab},H \} =0 ~~\Rightarrow ~~~  R^c ~V^d \epsi_{abcd} = 0 \label{DCsecondary2}
\ena
Note that to derive (\ref{DCsecondary2}) we did not need the identity (\ref{identity1}).

The conditions (\ref{DCsecondary1}), (\ref{DCsecondary2}) are the same as those derived in Sect. 9. , and likewise the 
constraint algebra is the same.

The doubly covariant formalism can be applied to geometric theories with a Lagrangian $d$-form $L=L(\phi, R)$ invariant under local gauge tangent space symmetries, and where the variation of the ``velocities" (i.e. curvatures) $R$ is given by $\delta R = \Dcal (\delta \phi)$, where $\Dcal$ is the (Lorentz) covariant derivative. Indeed consider the variational principle applied to the action
\eq
S= \int_{\Mcal^d} L (\phi_i, R_i)
\en
yielding  
\eq
\delta S =  \int_{\Mcal^d} \delta \phi_i { \dright L \over \partial \phi_i} + \Dcal (\delta \phi_i) { \dright L \over \partial R_i }=0
\en
and leading to the Euler-Lagrange equations:
\eq
\Dcal ~ { \dright L \over \partial R_i} - (-)^{p_i} { \dright L \over \partial \phi_i} =0 \label{ELeqsDC}
\en
Defining the momenta
\eq
\pi^i \equiv {\dright L \over \partial R_i} \label{momentadefDC}
\en
the $d$-form Hamiltonian density
\eq
H \equiv R_i ~\pi^i  - L \label{formHDC}
\en
does not depend on the ``velocities" $R_i$ since
\eq
{\dright H \over \partial R_i } = \pi^i - {\dright L \over \partial R_i}= 0 
\en
Thus $H$ depends on the $\phi_i$ and $\pi^i$:
\eq
H=H(\phi_i,\pi^i)
\en
and the form-analogue of the Hamilton equations reads:
\eq
R_i = (-1)^{(d+1)(p_i+1)} {\dright H \over \partial \pi^i} ,~~~\Dcal \pi^i =  (-)^{p_i+1} ~{\dright H \over \partial \phi_i} \label{formHEDC}
\en
These equations are derived by the same reasoning used for eq.s (\ref{formHE}).

\sect{Conclusions}

We have extended the covariant hamiltonian approach of ref.s \cite{CCF1}-\cite{CCF5}  with a 
 form-Legendre transformation that leads to a
consistent definition of form-Poisson brackets. In the $d=3$ vielbein gravity case, 
form-Dirac brackets can be defined. The algorithmic procedure of \cite{SCHS} can be
generalized in this formalism, and is applied to find gauge generators for gravity in 
$d=3$ and $d=4$. Finally a ``doubly covariant" hamiltonian is used in $d=4$ gravity.

The formalism proposed here can be applied as it stands to supergravity theories, where $p$-forms abound.
It could be worthwhile to use it for superspace lagrangians with integral forms, see for ex. \cite{integralforms1,integralforms2}.
Also, it appears to be particularly suited to noncommutative generalizations of gravity along the lines of ref.s \cite{NCgravity1,NCgravity2},
where the twist is defined in form language. 

\section*{Acknowledgement}

This work has been partially supported by Universit\`a del Piemonte Orientale research funds.

\vfill\eject
\end{document}